\documentclass{JHEP3}
\usepackage{psfig}

\title{Geometric Origin of CP Violation in an
Extra-Dimensional Brane World}

\author{
David Dooling \\
McGill University \\
Montreal, Quebec, H3A 2T8, CANADA \\
E-mail:
\email{dooling@hep.physics.mcgill.ca}}
\author{
Damien A. Easson \\
McGill University \\
Montreal, Quebec, H3A 2T8, CANADA\\
E-mail:
\email{easson@hep.physics.mcgill.ca}}
\author{
Kyungsik Kang \\
Brown University \\
Providence, RI 02912, USA \\
E-mail: \email{kang@het.brown.edu}}

\abstract{
The fermion mass hierarchy and finding a predictive mechanism
of the flavor mixing parameters remain two of the least
understood puzzles facing particle physics today.
In this work, we demonstrate how the realization of the Dirac
algebra in the presence of two extra spatial
dimensions leads to complex fermion field profiles in
the extra dimensions.
Dimensionally reducing to four dimensions leads to 
complex quark mass matrices in such a fashion that CP
violation necessarily follows.
We also present the generalization of the Randall-Sundrum
scenario to the case of a multi-brane, six-dimensional
brane-world and discuss how multi-brane worlds may shed light on 
the generation index of the SM matter content.
}
\preprint{MCGILL 02-03 \\ BROWN-HET-1298 \\ {\tt hep-ph/0202206}}
\begin{document}
\section{Introduction}

A fundamental explanation of both the quark flavor-mixing matrix and the 
fermion masses and their hierarchical structure persist to be two of the
most challenging problems of particle physics today.
A predictive mechanism for fermion mass generation is currently
lacking.
After spontaneous symmetry breaking, the quark mass term in
the langrangian reads:
\begin{equation}
\mathcal{L}_{\mathnormal \mbox{mass} } \mathnormal =
\frac{v}{\sqrt{2}} \left( \overline{u_{L_{i}}} h_{ij}^{(u)} u_{R_{j}} +
 \overline{d_{L_{i}}} h_{ij}^{(d)} d_{R_{j}} \right) + \mbox{h.c.}
\end{equation}
where the $h_{ij}$ are arbitrary $3 \times 3$ complex Yukawa
coupling matrices.
Within the standard model (SM) the fermion masses, the quark 
flavor-mixing angles and the CP violating phase are free parameters and
no relation exists among them.
The SM can accommodate the observed mass spectrum of the fermions but
unfortunately does not predict it.
Thus the calculability of the fermion masses remains an outstanding
theoretical challenge.
Our hope is that some predictive mechanism of fermion
mass generation exists and will place the understanding of fermion mass on a 
par with that of gauge boson mass.

The number of free parameters in the Yukawa sector eliminates any real
predictive power of this sector of the SM.
A first step one may take is to modify the Yukawa sector
in such a way that the four quark flavor-mixing parameters depend
solely on the quark masses themselves.
As an attempt to derive a relationship between the quark masses and
flavor-mixing parameters, mass matrix ans\"{a}tze based on 
flavor democracy with a suitable breaking so as to allow mixing 
between the quarks of nearest kinship
via nearest neighbor interactions was suggested about two decades
ago \cite{Weinberg:1977hb,Wilczek:1977uh,Rothman:1979ft,Kang:1981yg,Fritzsch:1978vd,Fritzsch:1979zq,DeRujula:1977ry,Georgi:1979dq}.
These early attempts are the first examples of ``strict calculability'';
i.e., mass matrices such that all flavor-mixing parameters 
depend solely on, and are determined by, the quark masses.
But the simple symmetric NNI texture leads to the experimentally
violated inequality $M_{top} < 110$ GeV, prompting
consideration of a less restricted form for the mass matrices
so as to retain calculability, yet be consistent
with experiment \cite{Kang:1997uv}.

After implementing this first step towards
gaining a deeper understanding of the Yukawa sector of the SM in the
guise of calculability, one may then attempt to
 confront the fundamental problem of explaining
the fermion mass hierarchy itself.
In this paper, we will address neither of the above issues, but
instead shall retreat even further into a simpler domain of the 
overall problem.
The quark mixing matrix contains four physical 
parameters, the three mixing angles and the single CP violating
phase of the Cabibo-Kobayashi-Maskawa (CKM) matrix.
Here we conjecture that the CP phase
in the quark flavor-mixing matrix may be a result of the existence of 
extra dimensions and the Dirac algebra realized in the presence
of these extra dimensions.
The notion of CP violation arising from the presence of extra
dimensions is not new, but was studied long ago
 \cite{Thirring:1972de}, and more recently in
\cite{Casadio:2001fe,Chang:2001uk,Chang:2001yn,Huang:2001np,Branco:2000rb,Sakamura:2000ik,Sakamura:1999fa}.

Our work is largely inspired by the papers \cite{Hung:2001hw,Huber:2000ie,Arkani-Hamed:1999dc,Mouslopoulos:2001uc,Kaplanet:2001} in which the fermion
mass and mixing hierarchies have been addressed within the context
of large extra dimensions (LED), both in the five- and six-dimensional
cases, as well as within the context of a single extra dimension
with warped geometry.
This paper addresses the same problem using two
extra dimensions with warped geometry, so as to explain the
existence of the CP phase in the quark-flavor mixing matrix.
Six dimensional extensions of the RS scenario have been
studied in \cite{Chodos:1999zt,Gherghettaet:2000,Collins:2001ni,Kogan:2001yr,Burgess:2001bn,Kim:2001rm}.

Another major theoretical problem in physics is that
of the hierarchy between the Planck scale 
(the rest mass of a flea) and particle physics scales,
such as the masses of the $W$ and $Z$ bosons.
String/M-theory seems to be the most promising candidate to form
a tight conceptual connection between gravity and particle
physics, with its attendant extra dimensions.
Some novel ideas concerning solutions to the gauge
hierarchy problem are rooted in the possibility that
the hierarchy is controlled by exotic features of the extra 
dimensions; namely, either they are very large and so the
hierarchy is generated by the volume of the extra
dimensions \cite{Arkani-Hamed:1998rs,Arkani-Hamed:1998nn,
Antoniadis:1998ig} or that the extra dimensions are warped, with the 
hierarchy generated by an exponential damping
\cite{Gogberashvili:1998vx,Randall:1999ee,Randall:1999vf}.
Here we explore
the implications of extra dimensions for the fermion
mass and mixing problem.
This investigation sheds light on the source of CP
violation. 

The organization of this paper is as follows.
We briefly review the Randall-Sundrum scenario and introduce
its extension to two extra dimensions.
We then introduce fundamental six-dimensional fermions
(eight-component objects), as well as a fundamental
Higgs scalar.
We present the equations for the fermion
zero modes, as well as the boundary conditions.
Previous treatments of the fermion boundary
conditions in six dimensions differ from the ones
presented here.
We solve for the Higgs zero mode and show,
for particular values of its six-dimensional
mass, how its profile is peaked away from the
six dimensional analogue of the hidden brane in the
brane set-up to be described.
The possibility of bulk SM fields within the RS scenario
has been extensively studied in \cite{Chang:1999nh}.
We demonstrate that the presence of two extra dimensions leads to
complex fermion field profiles in the extra dimension and 
show how this leads to CP violation.
We then conclude and briefly discuss some future avenues to be
investigated.

\section{Many-Brane, Six-Dimensional Extension of the 
Randall-Sundrum Solution}

Here we present a simple generalization of the RS scenario to the six
dimensional, multi-brane case.
In what follows, we consider one fundamental cell of the 
brane lattice (to be described below), but for completeness we
write down the entire brane-lattice solution.
Ultimately, one wants to understand not only the fermion mass
hierarchy and the flavor mixing parameters, but also the very
existence of the three fermion generations of the SM.
An interesting observation of Kogan et al \cite{Kogan:2001wp} is that
in multi-brane worlds, there exist ultra-light localized
and strongly coupled bulk fermion KK modes.
This leads to the possibility that for a 
given fundamental bulk 
fermion field with given SM gauge group transformation
properties, the generation index may be associated with KK
mode number, so that ultimately there is only one six-dimensional
species of up-type quarks, for example, and that the generation
structure is just a reflection of the existence of ultra-light
KK modes arising from the brane set-up and geometry of the
extra dimensions.
One stumbling block confronting this mode of 
understanding fermion generation structure is that the very
multi-brane set-up that gives rise to the family
structure also gives rise to the same number of ultralight
KK modes for all other fields, including the graviton.
This approach to the family index is currently under investigation,
though in the applications to follow we will consider only one 
fundamental cell of this brane crystal and hence will not
have any ultralight KK modes.

We now present a solution for the metric corresponding to a 
six dimensional, multi-brane extension of the RS scenario.
A five dimensional multi-brane extension is 
presented in \cite{Hatanaka:1999ac}.

To generalize to six dimensions, we consider a 
N $\times$ M lattice of N parallel 4-branes localized in the $\phi$ dimension 
orthogonal to M parallel 4-branes localized in the $\rho$ dimension.
3-Branes reside at their loci of intersection.
The action describing this set-up is given by the following three
terms:
\normalcolor
\begin{displaymath}
S = S_{gravity} + \sum_{i=1}^{N} S_{i} + \sum_{j=1}^{M} S_{j}
\end{displaymath}
where $S_{grav}$ is
\begin{displaymath}
S_{grav} = \int d^{4} x \int_{0}^{2 \pi} d \phi \int_{0}^{2 \pi} d
\rho \sqrt{-g} \left( \frac{1}{ \kappa_{6}^{2} } R - \sum_{i,j} 
\Lambda_{ij} \left[ \Theta \left( \phi - \phi_{i} \right) - \Theta
 \left( \phi - \phi_{i+1} \right) \right] \times \right.
\end{displaymath}
\begin{displaymath}
\left.
 \left[ \Theta \left( \rho
- \rho_{j} \right) - \Theta \left( \rho - \rho_{j+1} \right) \right] \right)
\end{displaymath}
and the terms in the action representing the 4-branes
are
\begin{displaymath}
S_{i} = - \int d^{4} x \int_{0}^{2 \pi} d \rho \sqrt{g^{\left( \phi = \phi_{i} \right)}} T_{\phi_{i}}
\end{displaymath}
and
\begin{displaymath}
S_{j} = -\int d^{4} x \int_{0}^{2 \pi} d \phi \sqrt{g^{\left( \rho = \rho_{j} \right)}} T_{\rho_{j}},
\end{displaymath}
where the $T_{\phi_{i}}$ are the tensions of the 4-branes
located at
 $\phi_{i}$ and $T_{\rho_{j}}$ are the tensions of the 4-branes
located at $\rho_{j}$. 
We are interested in the case of both extra dimensions being compact
and impose the following $S^{1}$ periodicity conditions on the 
extra coordinates:
\begin{displaymath}
\rho_{M+1} = 2 \pi
\end{displaymath}
\begin{displaymath}
\phi_{N+1} = 2 \pi
\end{displaymath}
Furthermore, we consider an orbifold
by imposing a pair of
$Z_{2}$ symmetries on the solution.
As in \cite{Hatanaka:1999ac}, we use the $S^{1}$ symmetry(ies) to 
define the position of the first set of 
 brane sources to be at the origin of the extra dimensions,
\begin{displaymath}
\phi_{1} = 0 = \rho_{1}
\end{displaymath}
In the above expressions for the 4-brane actions, 
the induced metric is
\begin{displaymath}
g_{ab}^{\rho = \rho_{j}} = g_{ab} \left( x^{\mu}, \phi, \rho = \rho_{j} \right)
\end{displaymath}
for the 4-branes localized in the $\phi$ direction and
\begin{displaymath}
g_{\alpha \beta}^{\phi = \phi_{i}} = g_{\alpha \beta} \left( x^{\mu}, \phi = 
\phi_{i}, \rho \right)
\end{displaymath}
for the 4-branes localized in the $\rho$ direction.
The six dimensional Einstein equations are
\begin{displaymath}
R_{N}^{M} - \frac{1}{2} \delta_{N}^{M} R = \frac{\kappa_{6}^{2}}{2} T_{N}^{M}
\end{displaymath}
where
\begin{eqnarray*}
T_{N}^{M} & = & -\sum_{i,j}^{N,M} \Lambda_{ij} \left[ \Theta \left(
\phi - \phi_{i} \right) - \Theta \left( \phi - \phi_{i+1} \right) 
\right] \left[ \Theta \left( \rho - \rho_{j} \right) - \Theta
 \left( \rho - \rho_{j+1} \right) \right] \delta_{N}^{M} \\
& = & -\sum_{i}^{N} \sqrt{\frac{-\mbox{det} g^{\phi = \phi_{i}}}{\mbox{det} g}} T_{\phi_{i}} \delta \left( \phi - \phi_{i} \right) \delta_{a}^{M} \delta^{a}_{N} \\
& = & -\sum_{j=1}^{M} \sqrt{\frac{-\mbox{det} g^{\rho = \rho_{j}}}{\mbox{det} g}} T_{\rho_{j}} \delta \left( \rho - \rho_{j} \right) \delta_{\alpha}^{M} \delta_{N}^{\alpha}
\end{eqnarray*}

We are not addressing any cosmological issues in this work, and for 
simplicity consider the following
static ans\"{a}tze for the metric:
\normalcolor
\begin{displaymath}
ds^{2} = A^{2} \left( \phi, \rho \right) \eta_{\mu \nu} dx^{\mu}
dx^{\nu} - B^{2} \left( \phi, \rho \right) d \phi^{2} - C^{2} \left(
\phi, \rho \right) d \rho^{2}
\end{displaymath}

With this ans\"{a}tz, the left-hand side of the Einstein 
equations are given by the following expressions, where dots
denote derivatives with respect to the $\rho$ coordinate and
primes denote derivatives with respect to the $\phi$ coordinate:
\begin{eqnarray*}
G_{\nu}^{\mu} & = & \frac{2}{A^{2}} \left[ \left(
\frac{\dot{A}}{A} \right)^{2} + \left( \frac{A^{\prime}}{A} \right)^{2} 
+ 2 \left( \frac{\ddot{A}}{A} + \frac{A^{\prime \prime}}{A} \right)
\right] \delta_{\nu}^{\mu} \\
G_{{\phi}}^{{\phi}} & = & \frac{2}{A^{2}} \left[ 5 \left( \frac{A^{\prime}}{A} \right)^{2} + 2 \frac{\ddot{A}}{A} + \left(
\frac{\dot{A}}{A} \right)^{2} \right] \\
\end{eqnarray*}
\begin{eqnarray*}
G_{{\rho}}^{{\rho}} & =  & \frac{2}{A^{2}} \left[ 5 \left(
\frac{\dot{A}}{A} \right)^{2} + 2 \frac{ A^{\prime \prime}}{A} + 
\left( \frac{A^{\prime}}{A} \right)^{2} \right] \\
G_{{\rho}}^{{\phi}} & = & \frac{4}{A^{4}} \left[
-A\dot{A^{\prime}} + 2 \dot{A} A^{\prime} \right] \\
G_{{\phi}}^{{\rho}} & = & G_{{\rho}}^{{\phi}}
\end{eqnarray*}
Taking the warp factor to be
\normalcolor

\begin{displaymath}
A = \frac{1}{e^{\sigma \left( \phi \right)} +
 e^{\gamma \left( \rho \right)}
 + 1}
\end{displaymath}
and
\begin{displaymath}
B \left( \phi, \rho \right) = A \left( \phi, \rho \right) e^{ \sigma \left( \phi \right)}
\end{displaymath}
\begin{displaymath}
C \left( \phi, \rho \right) = A \left( \phi, \rho \right) e^{ \gamma \left( \rho \right)}
\end{displaymath}
one can easily check that the nondiagonal elements of the Einstein tensor
vanish,
$G_{{\rho}}^{{\phi}} = 0$, and thus the ${\phi} - 
{\rho}$ component of the Einstein equations are trivially 
satisfied.

The remaining equations will be satisfied if the following relations
are fulfilled:
\normalcolor
\begin{displaymath}
10 \left( k_{\phi_{i}}^{2} + k_{\rho_{j}}^{2} \right) =
-\frac{\kappa_{6}^{2}}{2} \sum_{i,j} \Lambda_{ij} \left[ \Theta \left(
\phi - \phi_{i} \right) - \Theta \left( \phi - \phi_{i+1} \right)
\right] \times
\end{displaymath}
\begin{displaymath}
 \left[ \Theta \left( \rho - \rho_{j} \right) - \Theta \left(
\rho - \rho_{j} \right) \right],
\end{displaymath}
\begin{displaymath}
8 \left( k_{\rho_{j}} - k_{\rho_{j-1}} \right) = \frac{\kappa_{6}^{2}}{2}
T_{\rho_{j}},
\end{displaymath}
\begin{displaymath}
8 \left( k_{\phi_{i}} - k_{\phi_{i-1}} \right) = \frac{\kappa_{6}^{2}}{2}
T_{\phi_{i}},
\end{displaymath}
So we have
\normalcolor
\begin{displaymath}
\sigma \left( \phi \right) = 
k_{\phi_{1}} | \phi - \phi_{1} | \Theta \left( \phi - \phi_{1} \right) + 
\left( k_{\phi_{2}} - k_{\phi_{1}} \right) | \phi - \phi_{2} | \Theta \left( \phi - \phi_{2} \right) +
\end{displaymath}
\begin{displaymath}
 \left( k_{\phi_{3}} - k_{\phi_{2}} \right) | \phi
- \phi_{3} | \Theta \left( \phi - \phi_{3} \right) + \cdots
 +
\left( k_{\phi_{N}} - k_{\phi_{N-1}} \right) | \phi - \phi_{N}| 
\Theta \left( \phi - \phi_{N} \right)
\end{displaymath}
\begin{displaymath}
\gamma \left( \rho \right) =
k_{\rho_{1}} | \rho - \rho_{1} | \Theta \left( \rho - \rho_{1} \right) +
\left( k_{\rho_{2}} - k_{\rho_{1}} \right) | \rho - \rho_{2} | \Theta
\left( \rho - \rho_{2} \right) +
\end{displaymath}
\begin{displaymath}
 \left( k_{\rho_{3}} - k_{\rho_{2}} \right)
| \rho - \rho_{3} | \Theta \left( \rho - \rho_{3} \right) + \cdots
+ \left( k_{\rho_{M}} - k_{\rho_{M-1}} \right) |\rho - \rho_{M} |
\Theta \left( \rho - \rho_{M} \right)
\end{displaymath}
and
\normalcolor
\begin{displaymath}
k_{i} - k_{i-1} = \frac{\kappa_{6}^{2}}{16} T_{i}
\end{displaymath}
\begin{displaymath}
\kappa_{6}^{2} = \frac{ 16 \pi}{M_{6}^{4}}
\end{displaymath}
We 
work in units where $M_{6}^{-4} = 1$, so that
\normalcolor
\begin{displaymath}
k_{i} - k_{i-1} = \pi T_{i}
\end{displaymath}
and
for simplicity we take all the bulk cosmological constants to be equal
and
the magnitudes of the brane tensions to be the same:
$| T_{\phi_{i}} | = |T_{\rho_{j}}| = T$
Thus we have the relations

\begin{displaymath}
10 \left( 2 \right) k_{1}^{2} = -\frac{ 16 \pi}{\left( 2 \right) M_{6}^{4}} \Lambda
\end{displaymath}
\begin{displaymath}
k_{1} = \sqrt{ -\frac{2 \pi \Lambda}{5}} = k_{\phi_{1}} = k_{\rho_{1}}
\end{displaymath}
 Our expressions for the functions
$\sigma \left( \phi \right)$ and $\gamma \left( \rho \right)$
then become 
\begin{displaymath}
\sigma \left( \phi \right) = \sqrt{-\frac{2 \pi \Lambda}{5}} \phi \Theta 
\left( \phi \right) + \pi T \left[ -\left( \phi - \phi_{2} \right) 
\Theta \left( \phi - \phi_{2}\right) \right.
\end{displaymath}
\begin{displaymath}
+ \left( \phi - \phi_{3} \right) \Theta \left( \phi - \phi_{3} \right) - 
\left( \phi - \phi_{4} \right) \Theta \left( \phi - \phi_{4} \right)
\end{displaymath}
\begin{displaymath}
+ \left( \phi - \phi_{5} \right) \Theta \left( \phi - \phi_{5} \right) -
\left( \phi - \phi_{6} \right) \Theta \left( \phi - \phi_{6} \right)
\end{displaymath}
\begin{displaymath}
\left. + \ldots
 - \left( \phi - \phi_{N} \right) \Theta \left( \phi - \phi_{N}
\right) \right]
\end{displaymath}
\begin{displaymath}
\sigma \left( \phi_{c} \right) = \sigma \left( 0 \right) = 0 \rightarrow
\end{displaymath}
\begin{displaymath}
\phi_{c} = \frac{ \pi T \left( \phi_{2} - \phi_{3} + \phi_{4} - \phi_{5} +
\phi_{6} - \ldots + \phi_{N} \right)}{\left( \pi T - \sqrt{
-\frac{2 \pi \Lambda}{5}} \right)}
\end{displaymath}
and similarly,
\normalcolor
\begin{displaymath}
\gamma \left( \rho \right) = \sqrt{ -\frac{ 2 \pi \Lambda}{5}} \rho
\Theta \left( \rho \right) + \pi T \left[ -\left( \rho - \rho_{2} \right)
\Theta \left( \rho - \rho_{2} \right) \right.
\end{displaymath}
\begin{displaymath}
\left. + \ldots - \left( \rho - \rho_{M} \right) \Theta
\left( \rho - \rho_{M} \right) \right]
\end{displaymath}
\begin{displaymath}
\gamma \left( \rho_{c} \right) = \gamma \left( 0 \right) = 0 \rightarrow
\end{displaymath}
\begin{displaymath}
\rho_{c} = \frac{ \pi T \left( \rho_{2} - \ldots + \rho_{M} \right)}{
\left( \pi T - \sqrt{-\frac{ 2 \pi \Lambda}{5}} \right)}.
\end{displaymath}

\section{Six Dimensional Dirac Fermions}

In the seminal work of \cite{Arkani-Hamed:1999dc}, a framework is introduced for 
understanding both the fermion mass hierarchy and proton
stability without recourse to flavor symmetries 
in terms of higher dimensional geography.
Additional physics must be assumed in order to localize the fermions in
the flat extra dimension, but once this additional
scalar field is introduced, any coupling between chiral fermions
is exponentially suppressed because the two fields are
separated in space.
A key observation made by Huber and Shafi in \cite{Huber:2000ie} is that one
can get this exponential damping automatically from a 
non-factorizable geometry and that there is no need to assume 
additional physics.

 In both \cite{Huber:2000ie} and 
\cite{Arkani-Hamed:1999dc}, the effective four dimensional masses arise
from integrating over the extra dimensional Yukawa interactions
between the five dimensional Higgs and chiral fermion
fields.
The resulting four dimensional Yukawa coupling matrices
exhibit phenomenologically acceptable spectrums and mixing
(excluding CP violation) because of how these overlap
integrals can be tuned depending on the values of the
five dimensional mass term for the fermions in the action.
In the flat space scenario of \cite{Arkani-Hamed:1999dc}, the five dimensional mass
term serves to translate the gaussian profile of the fermions
along the extra dimension, so that the overlap integral of two
chiral fermions and the flat zero mode of the Higgs is itself
a gaussian, thus generating exponentially small effective
four dimensional Yukawa couplings.
In the warped space scenario of \cite{Huber:2000ie},
 with the $\frac{S^{1}}{Z_{2}}$
geometry of Randall-Sundrum, the right-handed zero modes
are peaked at the origin on the positive tension 3-brane, while
the left-handed zero modes are peaked at the other orbifold
fixed point around the negative tension 3-brane.
In this case, varying the value of the five dimensional mass
does not translate the fermion field profiles along the 
extra dimension; the right-handed and left-handed fermions remain
localized around the positive and negative tension 3-branes,
respectively.
What does change, however, is the width of the profile.
Hence, just as in the flat space scenario, small changes in the 
values of the five dimensional mass parameters lead to 
greatly amplified changes in the resulting four dimensional
Yukawa coupling matrices.

In both the flat and warped space scenarios in five dimensions,
the profiles for all fields in the extra dimension are real
and CP violation cannot be realized in a natural way.
While CP violation is not addressed in
\cite{Huber:2000ie}, a numerical
example is given where nine input parameters, essentially the ratios
of the five dimensional masses for the fermions appearing
in the five dimensional action to the AdS curvature scale, 
result in very good agreement with the quark mass spectrum and
absolute values of the elements of the $V_{CKM}$ matrix.
The agreement is striking, with just enough
disagreement in the mixing matrix to generate the suspicion 
that inclusion of CP violation may somehow bring about
even closer agreement between this type of model's predictions
and experiment.

The action for a fermion in the six dimensional background
considered here is~\cite{Grossman:1999ra}

\begin{equation} 
S = \int d^{4}x \int d \phi d \rho \sqrt{G} \left\{ E^{A}_{\alpha} \left[ \frac{i}{2} \overline{\psi} \gamma^{\alpha} \left( \stackrel{\rightarrow}{\partial_{A}} - \stackrel{\leftarrow}{\partial_{A}} \right) \psi \right] -m \left( \phi, \rho \right) \overline{\psi} \psi \right\}
\end{equation}
\begin{equation}\label{metric}
ds^{2} = A^{2} \left( \phi, \rho \right) \eta_{\mu \nu} dx^{\mu} dx^{\nu} -
B^{2} \left( \phi, \rho \right) d \phi^{2} - C^{2} \left( \phi, \rho \right)^{2}
\end{equation}
\begin{equation}
A = \frac{1}{e^{\sigma} + e^{ \gamma} -1}
\end{equation}
\begin{equation}
B = e^{\sigma} A
\end{equation}
\begin{equation}
C = e^{\gamma} A
\end{equation}
\begin{equation}
\sqrt{g} = \frac{ e^{\sigma} e^{\gamma} }{\left( e^{\sigma} + e^{\gamma} - 1\right)^{6}}
\end{equation}

We now show that the introduction
of another extra dimension leads to a natural explanation of 
CP violation.
In six spacetime dimensions the Dirac algebra is minimally
realized by $8 \times 8$ matrices.
A particularly convenient representation is that of~\cite{Hung:2001hw},
in which the ideas of \cite{Arkani-Hamed:1999dc}  have been extended to the six dimensional case.  This provides a theoretical motivation for
the so-called democratic mass matrices that have served as the 
starting point for many flavor symmetry approaches to the
quark mass hierarchy problem.
This convenient representation is presented below, where the
$\gamma_{5}$ in $\Gamma_{\phi}$ is the same $\gamma_{5}$
constructed from the four dimensional $\gamma$'s.

\begin{equation}
\Gamma_{0} = \left( \begin{array}{cc}
0 & +i \gamma_{0} \\
-i \gamma^{0} & 0  
\end{array} \right)
\end{equation}
\begin{equation}
\Gamma_{1} = \left( \begin{array}{cc}
0 & +i \gamma_{1} \\
-i \gamma^{1} & 0  
\end{array} \right)
\end{equation}
\begin{equation}
\Gamma_{2} = \left( \begin{array}{cc}
0 & +i \gamma_{2} \\
-i \gamma^{2} & 0  
\end{array} \right)
\end{equation}
\begin{equation}
\Gamma_{3} = \left( \begin{array}{cc}
0 & +i \gamma_{3} \\
-i \gamma^{3} & 0  
\end{array} \right)
\end{equation}
\begin{equation}
\Gamma_{\phi} = \left( \begin{array}{cc}
0 &  \gamma_{5} \\
 -\gamma^{5} & 0  
\end{array} \right)
\end{equation}
\begin{equation}
\Gamma_{\rho} = \left( \begin{array}{cc}
0 & +i_{4 \times 4} \\
+i_{4 \times 4} & 0  
\end{array} \right)
\end{equation}
\begin{equation}
\Gamma_{7} = \left( \begin{array}{cc}
1 & 0 \\
0 & -1  
\end{array} \right)
\end{equation}
\begin{equation}
\gamma_{0} = \left( \begin{array}{cccc}
0 & 0 & 1 & 0 \\
0 & 0 & 0 & 1 \\
1 & 0 & 0 & 0 \\
0 & 1 & 0 & 0 \end{array} \right)
\end{equation}
\begin{equation}
\gamma_{1} = \left( \begin{array}{cccc}
0 & 0 & 0 & -1 \\
0 & 0 & -1 & 0 \\
0 & 1 & 0 & 0 \\
1 & 0 & 0 & 0 \end{array} \right)
\end{equation}
\begin{equation}
\gamma_{2} = \left( \begin{array}{cccc}
0 & 0 & 0 & i \\
0 & 0 & -i & 0 \\
0 & -i & 0 & 0 \\
i & 0 & 0 & 0 \end{array} \right)
\end{equation}
\begin{equation}
\gamma_{3} = \left( \begin{array}{cccc}
0 & 0 & -1 & 0 \\
0 & 0 & 0 & 1 \\
1 & 0 & 0 & 0 \\
0 & -1 & 0 & 0 \end{array} \right)
\end{equation}
\begin{equation}
\gamma_{5} = \left( \begin{array}{cccc}
1 & 0 & 0 & 0 \\
0 & 1 & 0 & 0 \\
0 & 0 & -1 & 0 \\
0 & 0 & 0 & -1 \end{array} \right)
\end{equation}

Both the $\Gamma$ 's and the $\gamma$'s have the mostly minus signature,
accounting for the discrepancy between this representation and the
one
presented in \cite{Hung:2001hw}.

Using $\Gamma_{7}$, we can construct  projection operators
in the usual way to
get two four-component objects we call $\psi_{+}$ and $\psi_{-}$.
This procedure is
in perfect analogy with what is done in four dimensions 
(expressing a four dimensional fermion field in terms of its
left and right-handed components).

\begin{equation}
\psi_{+} = \frac{1}{2} \left( 1 - \Gamma_{7} \right) \psi
\end{equation}
\begin{equation}
\psi_{-} = \frac{1}{2} \left( 1 + \Gamma_{7} \right) \psi
\end{equation}
\begin{equation}
\psi = \psi_{+} + \psi_{-}
\end{equation}

Note that in this representation 
the six dimensional lorentz invariant fermion bilinear
$\overline{\psi} \psi$ has a complex coefficient when
expressed in terms of the four component fields $\psi_{+}$ and
$\psi_{-}$ and their four component conjugate fields
$\overline{\psi_{+}}$ and $\overline{\psi_{-}}$.
This leads to a real six dimensional mass
term which is complex when expressed in terms of the 
four component projections $\psi_{+}$ and $\psi_{-}$ (the 
six dimensional analogs of $\psi_{L}$ and $\psi_{R}$).

\begin{equation}
\overline{\psi} = \psi^{\dagger} \Gamma_{0} = \left( \psi_{+}^{\dagger}, \psi_{-}^{\dagger} \right) \left( \begin{array}{cc}
0 & i \gamma_{0} \\
-i \gamma_{0} & 0 \end{array} \right) = \left( -i \overline{\psi_{-}} , i \overline{\psi_{+}} \right)
\end{equation}

As in \cite{Mouslopoulos:2001uc},
we assume that the higher dimensional fermion mass term is the 
result of the coupling of the fermions with a scalar field
which has a nontrivial stable vacuum.
We assume this VEV to have a multi-kink solution, which we can 
express in terms of the functions $\sigma \left( \phi \right)$
and $\gamma \left( \rho \right)$.

After integrating by parts, we can recast the action as:

\begin{eqnarray*}
S & =  & \int \mbox{d}^4{x} \int \mbox{d} \phi \mbox{d} \rho \left(
\frac{i e^{\sigma} e^{\gamma} }{\left( e^{\sigma} + e^{\gamma} - 1 \right)^{5}}
\left[
\overline{\psi_{R+}} \gamma^{\mu} \partial_{\mu} \psi_{R+} + 
\overline{\psi_{L+}} \gamma^{\mu} \partial_{\mu} \psi_{L+} +
\overline{\psi_{R-}} \gamma^{\mu} \partial_{\mu} \psi_{R-} +
\overline{\psi_{L-}} \gamma^{\mu} \partial_{\mu} \psi_{L-} \right] \right. \\
 & & -\frac{1}{2} \overline{\psi_{R-}} \left( 
\frac{e^{\gamma}}{ \left( e^{\sigma} + e^{\gamma} - 1 \right)^{5}} 
\partial_{\phi} + \partial_{\phi} \frac{e^{\gamma}}{ \left( e^{\sigma} +
 e^{\gamma} - 1 \right)^{5}} \right) \psi_{L-} \\
 & &  +\frac{1}{2} \overline{\psi_{L-}} \left( 
\frac{e^{\gamma}}{ \left( e^{\sigma} + e^{\gamma} - 1 \right)^{5}} 
\partial_{\phi} + \partial_{\phi} \frac{e^{\gamma}}{ \left( e^{\sigma} +
 e^{\gamma} - 1 \right)^{5}} \right) \psi_{R-} \\
 & &  -\frac{1}{2} \overline{\psi_{R+}} \left( 
\frac{e^{\gamma}}{ \left( e^{\sigma} + e^{\gamma} - 1 \right)^{5}} 
\partial_{\phi} + \partial_{\phi} \frac{e^{\gamma}}{ \left( e^{\sigma} +
 e^{\gamma} - 1 \right)^{5}} \right) \psi_{L+} \\
 & & +\frac{1}{2} \overline{\psi_{L+}} \left( 
\frac{e^{\gamma}}{ \left( e^{\sigma} + e^{\gamma} - 1 \right)^{5}} 
\partial_{\phi} + \partial_{\phi} \frac{e^{\gamma}}{ \left( e^{\sigma} +
 e^{\gamma} - 1 \right)^{5}} \right) \psi_{R+} \\
 & & +\frac{i}{2} \overline{\psi_{R-}} \left( 
\frac{e^{\gamma}}{ \left( e^{\sigma} + e^{\gamma} - 1 \right)^{5}} 
\partial_{\rho} + \partial_{\rho} \frac{e^{\gamma}}{ \left( e^{\sigma} +
 e^{\gamma} - 1 \right)^{5}} \right) \psi_{L-} \\
 & &  +\frac{i}{2} \overline{\psi_{L-}} \left( 
\frac{e^{\gamma}}{ \left( e^{\sigma} + e^{\gamma} - 1 \right)^{5}} 
\partial_{\rho} + \partial_{\rho} \frac{e^{\gamma}}{ \left( e^{\sigma} +
 e^{\gamma} - 1 \right)^{5}} \right) \psi_{R-} \\
 & & -\frac{i}{2} \overline{\psi_{R+}} \left( 
\frac{e^{\gamma}}{ \left( e^{\sigma} + e^{\gamma} - 1 \right)^{5}} 
\partial_{\rho} + \partial_{\rho} \frac{e^{\gamma}}{ \left( e^{\sigma} +
 e^{\gamma} - 1 \right)^{5}} \right) \psi_{L+} \\
 & & -\frac{i}{2} \overline{\psi_{L+}} \left( 
\frac{e^{\gamma}}{ \left( e^{\sigma} + e^{\gamma} - 1 \right)^{5}} 
\partial_{\rho} + \partial_{\rho} \frac{e^{\gamma}}{ \left( e^{\sigma} +
 e^{\gamma} - 1 \right)^{5}} \right) \psi_{R+} \\
 & & \left. -m i \left( \frac{\sigma^{\prime}}{k} \right) 
 \left( \frac{\dot{\gamma}}{k} \right) 
\frac{i e^{\sigma} e^{\gamma} }{\left( e^{\sigma} + e^{\gamma} - 1 \right)^{5}}
\left( \overline{\psi_{L+}} \psi_{R-} + \overline{\psi_{R+}} \psi_{L-} -
\overline{\psi_{L-}} \psi_{R+} - \overline{\psi_{R-}} \psi_{L+} \right) \right)
\end{eqnarray*}

In order to extract the four dimensional physics, we 
write the action in terms of a sum over KK modes.
Ultimately, we will only be interested in the zero modes.

\begin{eqnarray*}
S & = & \sum_{m}  \sum_{n}
 \int \mbox{d}^{4} x  \left\{ i \overline{\psi_{n,m+}} \left( x 
\right)
\gamma^{\mu} \partial_{\mu} \psi_{n,m+} \left( x \right) + 
i \overline{\psi_{n,m-}} \left( x 
\right)
\gamma^{\mu} \partial_{\mu} \psi_{n,m-} \left( x \right) \right. \\
 & & \left.
 -m_{n,m+} \overline{\psi_{n,m+}} \left( x \right) \psi_{n,m+} \left( x \right)
 -m_{n,m-} \overline{\psi_{n,m-}} \left( x \right) \psi_{n,m-} \left( x \right)
 \right\}
\end{eqnarray*}

The decomposition of the KK modes is simplified when we
express
 $\psi_{R+}$ and $\psi_{L+}$ in the form:

\begin{equation}
\Psi_{(R,L)+} \left( x, \phi, \rho \right) = \sum_{m}  \sum_{n} 
\psi_{n,m+}^{R,L} \left( x \right) \left(
\frac{ e^{\sigma} e^{\gamma} }{ \left( e^{\sigma} + e^{\gamma} -1 \right)^{6}}
\right)^{-\frac{1}{2}} f_{n,m+}^{R,L} \left( \phi, \rho \right)
\end{equation}

and $\psi_{R-}$ and $\psi_{L-}$ in the form:

\begin{equation}
\Psi_{(R,L)-} \left( x, \phi, \rho \right) =  \sum_{m} \sum_{n} 
\psi_{n,m-}^{R,L} \left( x \right) \left(
\frac{ e^{\sigma} e^{\gamma} }{ \left( e^{\sigma} + e^{\gamma} -1 \right)^{6}}
\right)^{-\frac{1}{2}} f_{n,m-}^{R,L} \left( \phi, \rho \right)
\end{equation}

To reproduce the standard 4-d kinetic terms, we require the normalization
conditions

\begin{equation}
\int \sum_{k,j}  \sum_{n,m} \left( e^{\sigma} + e^{\gamma} - 1 \right) 
f_{n,k+}^{R^{\ast}} \left( \phi, \rho \right) f_{m,j+}^{R} \left(
\phi, \rho \right) \mbox{d} \phi \mbox{d} \rho = \delta_{mn}, \delta_{kj}
\end{equation}

\begin{equation}
\int \sum_{k,j}  \sum_{n,m} \left( e^{\sigma} + e^{\gamma} - 1 \right) 
f_{n,k-}^{R^{\ast}} \left( \phi, \rho \right) f_{m,j-}^{R} \left(
\phi, \rho \right) \mbox{d} \phi \mbox{d} \rho = \delta_{mn}, \delta_{kj}
\end{equation}

\begin{equation}
\int \sum_{k,j} \sum_{n,m} \left( e^{\sigma} + e^{\gamma} - 1 \right) 
f_{n,k+}^{L^{\ast}} \left( \phi, \rho \right) f_{m,j+}^{L} \left(
\phi, \rho \right) \mbox{d} \phi \mbox{d} \rho = \delta_{mn}, \delta_{kj}
\end{equation}

\begin{equation}
\int \sum_{k,j} \sum_{n,m} \left( e^{\sigma} + e^{\gamma} - 1 \right) 
f_{n,k-}^{L^{\ast}} \left( \phi, \rho \right) f_{m,j-}^{L} \left(
\phi, \rho \right) \mbox{d} \phi \mbox{d} \rho = \delta_{mn}, \delta_{kj}
\end{equation}

In order to read off the equations of motion that the $f$'s must solve,
we need to simplify some terms in the action.
For example,

\begin{eqnarray*}
\frac{1}{2} \overline{\psi_{L+}} 
\left( 
\frac{e^{\gamma}}{ \left( e^{\sigma} + e^{\gamma} - 1 \right)^{5}} 
\partial_{\phi} + \partial_{\phi} \frac{e^{\gamma}}{ \left( e^{\sigma} +
 e^{\gamma} - 1 \right)^{5}} \right) \psi_{R+} & = & \\
 -\frac{1}{2} \sum_{l,m=0}^{\infty} \overline{\psi_{L+l,m}} \left( x \right)
f_{L+l,m}^{\ast} \left( \phi, \rho \right) \frac{ \left( e^{\sigma} + 
e^{\rho} - 1 \right) e^{\gamma} \sigma^{\prime} }{ e^{\sigma} 
e^{\gamma} } \sum_{n,p=0}^{\infty} \psi_{R+n,p} \left( x \right) 
f_{R+n,p} \left( \phi, \rho \right) & & \\
+ 3 \sum_{l,m=0}^{\infty} \overline{\psi_{L+l,m}} \left( x \right) 
f_{L+l,m}^{\ast} \left( \phi, \rho \right) e^{\sigma} e^{\gamma} 
\sigma^{\prime} \sum_{n,p=0}^{\infty} \psi_{R+n,p} \left( x \right) f_{R+n,p}
\left( \phi, \rho \right) & & \\
+ \sum_{l,m=0}^{\infty} \overline{\psi_{L+l,m}} \left( x \right)
f_{L+l,m}^{\ast} \left( \phi, \rho \right) \frac{
\left( e^{\sigma} + e^{\gamma} - 1 \right)}{ e^{\sigma}} \sum_{n,p=0}^{\infty}
 \psi_{R+n,p} \left( x \right) f_{R+n,p}^{\prime} \left( \phi, \rho \right)
 & & \\
-\frac{5}{2} \sum_{l,m=0}^{\infty} \overline{\psi_{L+l,m}} \left( x \right)
f_{L+l,m}^{\ast} \left( \phi, \rho \right) \sigma^{\prime} 
\sum_{n,p=0}^{\infty} \psi_{R+n,p} \left( x \right) f_{R+n,p} 
\left( \phi, \rho \right) & &
\end{eqnarray*}
with similar expressions holding for the other terms.

We are interested in the equations for the zero modes
$( m_{n} = 0 )$,
found by means of varying $S$ with respect to $f_{R+0}^{\ast},
f_{L+0}^{\ast}, f_{L-0}^{\ast}, f_{R-0}^{\ast}$.

The normalization conditions reproduce the desired
 four dimensional kinetic energy terms, hence it is only necessary
to vary the remaining parts of $S$.
For example, the remaining
terms involving $f_{L+0}^{\ast}$ are:

\begin{eqnarray*}
-\frac{1}{2} \overline{\psi_{L+0}} \left( x \right) f_{L+0}^{\ast} 
\left( \phi, \rho \right) \frac{ \left( e^{\sigma} + e^{\gamma} 
-1 \right) e^{\gamma} }{ e^{\sigma} e^{\gamma} } \sigma^{\prime}
\psi_{R+0} \left( x \right) f_{R+0} \left( \phi, \rho \right) & & \\
+ 3 \overline{\psi_{L+0}} \left( x \right) f_{L+0}^{\ast} \left( \phi, 
\rho \right) e^{\sigma} e^{\gamma} \sigma^{\prime} \psi_{R+0} \left( x
\right) f_{R+0} \left( \phi, \rho \right) & & \\
+ \overline{\psi_{L+0}} \left( x \right) f_{L+0}^{\ast} \left( \phi, 
\rho \right) \frac{ \left( e^{\sigma} + e^{\gamma} - 1 \right) }{
e^{\sigma} } \psi_{R+0} \left( x \right) f_{R+0}^{\prime} \left(
\phi, \rho \right) & & \\
-\frac{5}{2} \overline{\psi_{L+0}} \left( x \right) f_{L+0}^{\ast}
\left( \phi, \rho \right) \sigma^{\prime} \psi_{R+0} \left( x \right)
f_{R+0} \left( \phi, \rho \right) & & \\
+\frac{i}{2} \overline{\psi_{L+0}} \left( x \right) f_{L+0}^{\ast} \left(
\phi, \rho \right) \frac{ \left( e^{\sigma} + e^{\gamma} - 1 \right) 2
 e^{\sigma} \dot{\gamma} }{ e^{\sigma} e^{\gamma} } \psi_{R+0} \left(
 x \right) f_{R+0} \left( \phi, \rho \right)  & & \\
 - 3 i \overline{\psi_{L+0}} \left( x \right) f_{L+0}^{\ast} \left(
\phi, \rho \right) e^{\gamma} e^{\sigma} \dot{\gamma} \psi_{R+0} \left(
x \right) f_{R+0} \left( \phi, \rho \right) & & \\
-i \overline{\psi_{L+0}} \left( x \right) f_{L+0}^{\ast} \left( \phi, 
\rho \right) \frac{ \left( e^{\sigma} + e^{\gamma} - 1 \right) }{
e^{\gamma} } \psi_{R+0} \left( x \right) \dot f_{R+0} \left( \phi,
 \rho \right) & & \\
+ \frac{i5}{2} \overline{\psi_{L+0}} \left( x \right) f_{L+0}^{\ast} \left(
\phi, \rho \right) \dot{\gamma} \psi_{R+0} \left( x \right) 
 f_{R+0} \left( \phi, \rho \right) & & \\
- i m \left( \frac{ \sigma^{\prime} }{ k } \right)
\left( \frac{ \dot{\gamma} }{ k } \right) \overline{\psi_{L+0}} \left(
x \right) f_{L+0}^{\ast} \left( \phi, \rho \right) \psi_{R-0} \left( x
 \right) f_{R-0} \left( \phi, \rho \right) & &
\end{eqnarray*}

Noting that the two four dimensional right-handed and 
left-handed spinors $\psi_{+}$ and $\psi_{-}$ are 
complex conjugates of each other 
\cite{Hung:2001hw}, we arrive at the 
following equation for the right handed zero mode
(recall that dots denote derivatives with respect to $\rho$ and
primes denote derivatives with respect to $\phi$):

\begin{eqnarray*}
-\frac{1}{2} \frac{ \left( e^{\sigma} + e^{\gamma} - 1 \right) }
{ e^{\sigma} e^{\gamma} } e^{\gamma} \sigma^{\prime} f_{R+0}
+ 3 e^{\sigma} e^{\gamma} \sigma^{\prime} f_{R+0} & & \\
 + \frac{ \left( e^{\sigma} + e^{\gamma} - 1 \right) }{ e^{\sigma} }
f_{R+0}^{\prime} - \frac{5}{2} \sigma^{\prime} f_{R+0} + \frac{i}{2}
 \frac{ \left( e^{\sigma} + e^{\gamma} - 1 \right) e^{\sigma} 
\dot{\gamma} }{ e^{\sigma} e^{\gamma} } f_{R+0} & & \\
- 3 i e^{\gamma} e^{\sigma} \dot{\gamma} f_{R+0} -
 i \frac{ \left( e^{\sigma} + e^{\gamma} - 1 \right) }{ e^{\gamma} }
 \dot{f}_{R+0} & & \\
+ i \frac{5}{2} \dot{\gamma} f_{R+0} - i m \left( \frac{
\sigma^{\prime} }{ k } \right) \left( \frac{ \dot{\gamma} }{ k } \right)
f_{R-0} & = & 0
\end{eqnarray*}

Similarly, we find the following remaining equations for the 
other zero modes:

\begin{eqnarray*}
-\frac{1}{2} \frac{ \left( e^{\sigma} + e^{\gamma} - 1 \right) }
{ e^{\sigma} e^{\gamma} } e^{\gamma} \sigma^{\prime} f_{R-0}
+ 3 e^{\sigma} e^{\gamma} \sigma{\prime} f_{R-0} & & \\
 + \frac{ \left( e^{\sigma} + e^{\gamma} - 1 \right) }{ e^{\sigma} }
f_{R-0}^{\prime} - \frac{5}{2} \sigma^{\prime} f_{R-0} - \frac{i}{2}
 \frac{ \left( e^{\sigma} + e^{\gamma} - 1 \right) e^{\sigma} 
\dot{\gamma} }{ e^{\sigma} e^{\gamma} } f_{R-0} & & \\
+ 3 i e^{\gamma} e^{\sigma} \dot{\gamma} f_{R-0} +
 i \frac{ \left( e^{\sigma} + e^{\gamma} - 1 \right) }{ e^{\gamma} }
 \dot f_{R-0} & & \\
- i \frac{5}{2} \dot{\gamma} f_{R-0} + i m \left( \frac{
\sigma^{\prime} }{ k } \right) \left( \frac{ \dot{\gamma} }{ k } \right)
f_{R+0} & = & 0
\end{eqnarray*}

\begin{eqnarray*}
+\frac{1}{2} \frac{ \left( e^{\sigma} + e^{\gamma} - 1 \right) }
{ e^{\sigma} e^{\gamma} } e^{\gamma} \sigma^{\prime} f_{L+0}
- 3 e^{\sigma} e^{\gamma} \sigma{\prime} f_{L+0} & & \\
 - \frac{ \left( e^{\sigma} + e^{\gamma} - 1 \right) }{ e^{\sigma} }
f_{L+0}^{\prime} + \frac{5}{2} \sigma^{\prime} f_{L+0} + \frac{i}{2}
 \frac{ \left( e^{\sigma} + e^{\gamma} - 1 \right) e^{\sigma} 
\dot{\gamma} }{ e^{\sigma} e^{\gamma} } f_{L+0} & & \\
- 3 i e^{\gamma} e^{\sigma} \dot{\gamma} f_{L+0} -
 i \frac{ \left( e^{\sigma} + e^{\gamma} - 1 \right) }{ e^{\gamma} }
 \dot f_{L+0} & & \\
+ i \frac{5}{2} \dot{\gamma} f_{L+0} - i m \left( \frac{
\sigma^{\prime} }{ k } \right) \left( \frac{ \dot{\gamma} }{ k } \right)
f_{L-0} & = & 0
\end{eqnarray*}

\begin{eqnarray*}
+\frac{1}{2} \frac{ \left( e^{\sigma} + e^{\gamma} - 1 \right) }
{ e^{\sigma} e^{\gamma} } e^{\gamma} \sigma^{\prime} f_{L-0}
- 3 e^{\sigma} e^{\gamma} \sigma^{\prime} f_{L-0} & & \\
 - \frac{ \left( e^{\sigma} + e^{\gamma} - 1 \right) }{ e^{\sigma} }
f_{L-0}^{\prime} + \frac{5}{2} \sigma^{\prime} f_{L-0} - \frac{i}{2}
 \frac{ \left( e^{\sigma} + e^{\gamma} - 1 \right) e^{\sigma} 
\dot{\gamma} }{ e^{\sigma} e^{\gamma} } f_{L-0} & & \\
+ 3 i e^{\gamma} e^{\sigma} \dot{\gamma} f_{L-0} +
 i \frac{ \left( e^{\sigma} + e^{\gamma} - 1 \right) }{ e^{\gamma} }
 \dot f_{L-0} & & \\
- i \frac{5}{2} \dot{\gamma} f_{L-0} + i m \left( \frac{
\sigma^{\prime} }{ k } \right) \left( \frac{ \dot{\gamma} }{ k } \right)
f_{L+0} & = & 0
\end{eqnarray*}

Because $f_{L+0}^{\ast} = f_{L-0}$, this is the complex conjugate
of the previous equation.

The equations for the right-handed zero modes are

\begin{eqnarray*}
-\frac{1}{2} \frac{ \left( e^{\sigma} + e^{\gamma} - 1 \right) }
{ e^{\sigma} e^{\gamma} } e^{\gamma} \sigma^{\prime} f_{R-} + 
3 e^{\sigma} e^{\gamma} \sigma^{\prime} f_{R-} + \frac{ \left(
e^{\sigma} + e^{\gamma} - 1 \right) }{ e^{\sigma} } f_{R-}^{\prime}
 & & \\
-\frac{5}{2} \sigma^{\prime} f_{R-} - i\frac{1}{2} \frac{ \left(
 e^{\sigma} + e^{\gamma} - 1 \right) }{ e^{\sigma} e^{\gamma} }
e^{\sigma} \dot{\gamma} f_{R-} + 3i e^{\gamma} e^{\sigma} 
\dot{\gamma} f_{R-} & & \\
+ i \frac{ \left( e^{\sigma} + e^{\gamma} - 1 \right) }{
e^{\gamma} } \dot{f_{R-}} - i \frac{5}{2} \dot{\gamma} f_{R-} + 
im \left( \frac{ \sigma^{\prime} }{k} \right)
 \left( \frac{ \dot{\gamma} }{ k } \right) f_{R+} & = & 0
\end{eqnarray*}

\begin{eqnarray*}
-\frac{1}{2} \frac{ \left( e^{\sigma} + e^{\gamma} - 1 \right) }
{ e^{\sigma} e^{\gamma} } e^{\gamma} \sigma^{\prime} f_{R+} + 
3 e^{\sigma} e^{\gamma} \sigma^{\prime} f_{R+} + \frac{ \left(
e^{\sigma} + e^{\gamma} - 1 \right) }{ e^{\sigma} } f_{R+}^{\prime}
 & & \\
-\frac{5}{2} \sigma^{\prime} f_{R+} + i\frac{1}{2} \frac{ \left(
 e^{\sigma} + e^{\gamma} - 1 \right) }{ e^{\sigma} e^{\gamma} }
e^{\sigma} \dot{\gamma} f_{R+} - 3i e^{\gamma} e^{\sigma} 
\dot{\gamma} f_{R+} & & \\
- i \frac{ \left( e^{\sigma} + e^{\gamma} - 1 \right) }{
e^{\gamma} } \dot{f_{R+}} + i \frac{5}{2} \dot{\gamma} f_{R+} - 
im \left( \frac{ \sigma^{\prime} }{k} \right)
 \left( \frac{ \dot{\gamma} }{ k } \right) f_{R-} & = & 0
\end{eqnarray*}

Because $f_{R+} = f_{R-}^{\ast}$, 
we are free
to write
\begin{eqnarray}
f_{R+}  & = & U + i V \\
f_{R-} & = & U - i V
\end{eqnarray}

where
$U$ and $V$ are real.
Setting the real and imaginary parts of this zero mode equation
to zero, we find 
equations:

\begin{eqnarray*}
-\frac{1}{2} \frac{ \left( e^{\sigma} + e^{\gamma} - 1 \right) }{
 e^{\sigma} e^{\gamma} } e^{\gamma} \sigma^{\prime} U_{R} + 
3 e^{\sigma} e^{\gamma} \sigma^{\prime} U_{R} + \frac{ \left(
 e^{\sigma} + e^{\gamma} - 1 \right) }{ e^{\sigma} } U^{\prime}_{R} 
 & & \\
 -\frac{5}{2} \sigma^{\prime} U_{R} - \frac{1}{2} \frac{ \left(
e^{\sigma} + e^{\gamma} - 1 \right) }{ e^{\sigma} e^{\gamma} }
e^{\sigma} \dot{\gamma} V_{R} + 3 e^{\gamma} e^{\sigma} \dot{\gamma} V_{R}
 + \frac{ \left( e^{\sigma} + e^{\gamma} - 1 \right) }{
e^{\gamma} } \dot{V}_{R} & & \\
-\frac{5}{2} \dot{\gamma} V_{R} - m \left( \frac{ \sigma^{\prime} }{
k} \right) \left( \frac{ \dot{\gamma} }{ k } \right) V_{R} & = & 0
\end{eqnarray*}

and
\begin{eqnarray*}
\frac{1}{2} \frac{ \left( e^{\sigma} + e^{\gamma} - 1 \right) }{
 e^{\sigma} e^{\gamma} } e^{\gamma} \sigma^{\prime} V_{R} - 
3 e^{\sigma} e^{\gamma} \sigma^{\prime} V_{R} - \frac{ \left(
 e^{\sigma} + e^{\gamma} - 1 \right) }{ e^{\sigma} } V^{\prime}_{R} 
 & & \\
 +\frac{5}{2} \sigma^{\prime} V_{R} - \frac{1}{2} \frac{ \left(
e^{\sigma} + e^{\gamma} - 1 \right) }{ e^{\sigma} e^{\gamma} }
e^{\sigma} \dot{\gamma} U_{R} + 3 e^{\gamma} e^{\sigma} \dot{\gamma} U_{R}
 + \frac{ \left( e^{\sigma} + e^{\gamma} - 1 \right) }{
e^{\gamma} } \dot{U}_{R} & & \\
-\frac{5}{2} \dot{\gamma} U_{R} + m \left( \frac{ \sigma^{\prime} }{
k} \right) \left( \frac{ \dot{\gamma} }{ k } \right) U_{R} & = & 0
\end{eqnarray*}

Similarly
setting the real and imaginary parts of the
left-handed 
 zero mode
equation to zero, we find:
\begin{eqnarray*}
\frac{1}{2} \frac{ \left( e^{\sigma} + e^{\gamma} - 1 \right) }{
 e^{\sigma} e^{\gamma} } e^{\gamma} \sigma^{\prime} U_{L} - 
3 e^{\sigma} e^{\gamma} \sigma^{\prime} U_{L} - \frac{ \left(
 e^{\sigma} + e^{\gamma} - 1 \right) }{ e^{\sigma} } U^{\prime}_{L} 
 & & \\
 +\frac{5}{2} \sigma^{\prime} U_{L} - \frac{1}{2} \frac{ \left(
e^{\sigma} + e^{\gamma} - 1 \right) }{ e^{\sigma} e^{\gamma} }
e^{\sigma} \dot{\gamma} V_{L} + 3 e^{\gamma} e^{\sigma} \dot{\gamma} V_{L}
 + \frac{ \left( e^{\sigma} + e^{\gamma} - 1 \right) }{
e^{\gamma} } \dot{V}_{L} & & \\
-\frac{5}{2} \dot{\gamma} V_{L} - m \left( \frac{ \sigma^{\prime} }{
k} \right) \left( \frac{ \dot{\gamma} }{ k } \right) V_{L} & = & 0
\end{eqnarray*}

and

\begin{eqnarray*}
-\frac{1}{2} \frac{ \left( e^{\sigma} + e^{\gamma} - 1 \right) }{
 e^{\sigma} e^{\gamma} } e^{\gamma} \sigma^{\prime} V_{L} + 
3 e^{\sigma} e^{\gamma} \sigma^{\prime} V_{L} + \frac{ \left(
 e^{\sigma} + e^{\gamma} - 1 \right) }{ e^{\sigma} } V^{\prime}_{L} 
 & & \\
 -\frac{5}{2} \sigma^{\prime} V_{L} - \frac{1}{2} \frac{ \left(
e^{\sigma} + e^{\gamma} - 1 \right) }{ e^{\sigma} e^{\gamma} }
e^{\sigma} \dot{\gamma} U_{L} + 3 e^{\gamma} e^{\sigma} \dot{\gamma} U_{L}
 + \frac{ \left( e^{\sigma} + e^{\gamma} - 1 \right) }{
e^{\gamma} } \dot{U}_{L} & & \\
-\frac{5}{2} \dot{\gamma} U_{L} + m \left( \frac{ \sigma^{\prime} }{
k} \right) \left( \frac{ \dot{\gamma} }{ k } \right) U_{L} & = & 0
\end{eqnarray*}

In order to find solutions that respect the symmetries
$\phi \rightarrow -\phi$ and $\rho \rightarrow -\rho$
of this brane background, we need to know how these two $Z_{2}$
symmetries are realized on fermions in this background:
\begin{equation}
\psi \left( x^{\mu}, \phi, \rho \right) 
\rightarrow \Psi \left( x^{\mu}, \phi, \rho \right) = \pm
\Gamma^{0} \Gamma^{1} \Gamma^{2} \Gamma^{3} \Gamma^{\rho}
\psi \left( x^{\mu}, \phi_{c} - \phi, \rho \right)
\end{equation}

and

\begin{equation}
\psi \left( x^{\mu}, \phi, \rho \right) 
\rightarrow \Psi \left( x^{\mu}, \phi, \rho \right) = \pm
\Gamma^{\rho} \Gamma^{7} 
\psi \left( x^{\mu}, \phi, \rho_{c} - \rho \right)
\end{equation}

Because only fermion bilinears appear in the action,
the choice of sign is arbritrary.
In order to derive definite boundary conditions,
we choose the signs given below.
One can easily show that
\begin{equation}
S_{\phi} = -\Gamma^{0} \Gamma^{1} \Gamma^{2} \Gamma^{3} \Gamma^{\rho}
\end{equation}
satisfies
\begin{equation}
S_{\phi}^{-1} \Gamma^{M} S_{\phi} 
= 
\Lambda_{N}^{M} \Gamma^{N}
\end{equation}
where $\Lambda$ corresponds to the Lorentz transformation
\begin{equation}
\Lambda = \left( \begin{array}{cccccc}
1 & 0 & 0 & 0 & 0 & 0 \\
0 & 1 & 0 & 0 & 0 & 0 \\
0 & 0 & 1 & 0 & 0 & 0 \\
0 & 0 & 0 & 1 & 0 & 0 \\
0 & 0 & 0 & 0 & -1 & 0 \\
0 & 0 & 0 & 0 & 0 & 1 
\end{array} \right)
\end{equation}
and 
\begin{equation}
L = i \overline{\psi} \Gamma^{N} \partial_{N} \psi
\end{equation}
is invariant.

Similarly,
\begin{equation}
S_{\rho} = -\Gamma_{\rho} \Gamma_{7}
\end{equation}
satisfies
\begin{equation}
S_{\rho}^{-1} \Gamma^{M} S_{\rho} =
\Lambda_{N}^{M} \Gamma^{N}
\end{equation}
where $\Lambda$ corresponds to the L.T.
\begin{equation}
\Lambda = \left( \begin{array}{cccccc}
1 & 0 & 0 & 0 & 0 & 0 \\
0 & 1 & 0 & 0 & 0 & 0 \\
0 & 0 & 1 & 0 & 0 & 0 \\
0 & 0 & 0 & 1 & 0 & 0 \\
0 & 0 & 0 & 0 & 1 & 0 \\
0 & 0 & 0 & 0 & 0 & -1 
\end{array} \right)
\end{equation}
and 
\begin{equation}
L = i \overline{\psi} \Gamma^{N} \partial_{N} \psi
\end{equation}
is invariant.

With $S_{\rho} = -\Gamma_{\rho} \Gamma_{7}$, this 
symmetry translates into
\begin{equation}
\left( \begin{array}{c}
\psi_{+}^{\prime} \\
\psi_{-}^{\prime} \end{array} \right)
= -\Gamma_{\rho} \Gamma_{7} \left( \begin{array}{c}
\psi_{+} \\
\psi_{-} \end{array}
\right)
\end{equation}

\begin{equation}
\left( \begin{array}{c}
\psi_{+}^{\prime} \\
\psi_{-}^{\prime} \end{array} \right)
= 
\left( \begin{array}{c}
+i \psi_{-}^{\prime} \\
-i \psi_{+}^{\prime} \end{array} \right)
\end{equation}

\begin{equation}
\psi_{+}^{\prime} \left( x^{\mu}, \phi, \rho \right) \rightarrow
\Psi_{+} \left( x^{\mu}, \phi, \rho \right) = +i \psi_{-}
\left( x^{\mu}, \phi, \rho_{c} - \rho \right) 
\end{equation}
\begin{equation}
\psi_{-}^{\prime} \left( x^{\mu}, \phi, \rho \right) \rightarrow
\Psi_{-} \left( x^{\mu}, \phi, \rho \right) = 
- i \psi_{+} \left( x^{\mu}, \phi, \rho_{c} - \rho \right)
\end{equation}
Our periodic boundary condition is
\begin{equation}
\psi_{\pm} \left( x^{\mu}, \phi, \rho \right) = 
\Psi_{\pm} \left( x^{\mu}, \rho_{c} + \rho \right)
\end{equation}
\begin{equation}
\psi_{\pm} \left( x^{\mu}, \phi, \rho \right) =
\psi_{\pm} \left( x^{\mu}, \phi, 2 \rho_{c} + \rho \right)
\end{equation}

\begin{eqnarray*}
\psi_{\pm} \left( x^{\mu}, \phi, -\rho \right) & = &
\Psi_{\pm} \left( x^{\mu}, \phi, \rho_{c} - \rho \right) \\
 & = & \pm i \psi_{\mp} \left( x^{\mu}, \phi, \rho \right)
\end{eqnarray*}

\begin{eqnarray*}
\psi_{\pm} \left( x^{\mu}, \phi, \rho_{c} + \rho \right) & = &
\Psi_{\pm} \left( x^{\mu}, \phi, \rho \right) \\
 & = & \pm i \psi_{\mp} \left( x^{\mu}, \phi,\rho_{c} - \rho \right)
\end{eqnarray*}
We immediately recognize $\rho = 0, \rho_{c}$ to be the
fixed points of the orbifold.

It is convenient to rewrite $\psi_{\pm}$ as
\begin{equation}
\psi_{\pm} = U \pm i V
\end{equation}
It then follows that

\begin{eqnarray*}
U \left( x^{\mu}, \phi, -\rho \right) & = & 
 V \left( x^{\mu}, \phi, \rho \right) \\
V \left( x^{\mu}, \phi, -\rho \right) & = & 
 U \left( x^{\mu}, \phi, \rho \right)
\end{eqnarray*}
and 

\begin{eqnarray*}
U \left( x^{\mu}, \phi, \rho + \rho_{c} \right) & = & 
 V \left( x^{\mu}, \phi, \rho_{c} - \rho \right) \\
V \left( x^{\mu}, \phi, \rho_{c} + \rho \right) & = & 
 U \left( x^{\mu}, \phi, \rho_{c} - \rho \right)
\end{eqnarray*}

With
\begin{equation}
S_{\phi} = - \Gamma^{0} \Gamma^{1} \Gamma^{2} 
\Gamma^{3} \Gamma^{\rho}
\end{equation}
the $\phi \rightarrow -\phi$ symmetry translates to
\begin{equation}
\psi_{\pm}^{\prime} \left( x^{\mu}, \phi, \rho \right) 
\rightarrow \Psi_{\pm} \left( x^{\mu}, \phi, \rho \right)
= -\gamma^{5} \psi_{\pm} \left( x^{\mu}, \phi_{c} - \phi,
\rho \right)
\end{equation}
Combine this symmetry with the periodic b.c.
\begin{eqnarray*}
\psi_{\pm} \left( x^{\mu}, \phi, \rho \right) & = & 
\Psi_{\pm} \left( x^{\mu}, \phi + \phi_{c}, \rho \right) \\
 & = & \psi_{\pm} \left( x^{\mu}, 2 \phi_{c} + \phi, \rho \right)
\end{eqnarray*}
and
\begin{eqnarray*}
\psi_{\pm} \left( x^{\mu}, -\phi, \rho \right) & = & 
\Psi_{\pm} \left( x^{\mu}, \phi_{c} - \phi, \rho \right) \\
 & = & -\gamma_{5} \psi_{\pm} \left( x^{\mu},
  \phi_{c} - \left( \phi_{c} - \phi \right), \rho \right)
\end{eqnarray*}
and
\begin{eqnarray*}
\psi_{\pm} \left( x^{\mu}, \phi + \phi_{c}, \rho \right) & = & 
\Psi_{\pm} \left( x^{\mu}, \phi, \rho \right) \\
 & = & -\gamma_{5} \psi_{\pm} \left( x^{\mu},
  \phi_{c} - \phi, \rho \right)
\end{eqnarray*}
This shows that $\phi = 0, \phi_{c}$ are fixed points.

One can subsequently define the chiral components
of $\psi_{+}, \psi_{-}$ by using the usual operators
\begin{equation}
P_{R,L} = \frac{ 1 \pm \gamma^{5} }{2}
\end{equation}
with
\begin{eqnarray*}
\psi_{+,R} & = & P_{R} \psi_{+} \\
\psi_{+,L} & = & P_{L} \psi_{+} \\
\psi_{-,R} & = & P_{R} \psi_{-} \\
\psi_{-,L} & = & P_{L} \psi_{-} 
\end{eqnarray*}
It follows that
\begin{eqnarray*}
U \left( x^{\mu}, -\phi, \rho \right) & = & 
-\gamma_{5} U \left( x^{\mu}, \phi, \rho \right) \\
V \left( x^{\mu}, -\phi, \rho \right) & = & 
-\gamma_{5} V \left( x^{\mu}, \phi, \rho \right) \\
U \left( x^{\mu}, \phi + \phi_{c}, \rho \right) & = & 
-\gamma_{5} U \left( x^{\mu}, \phi_{c} - \phi, \rho \right) \\
V \left( x^{\mu}, \phi + \phi_{c}, \rho \right) & = & 
-\gamma_{5} V \left( x^{\mu}, \phi - \phi_{c}, \rho \right)
\end{eqnarray*}

Summarizing all the b.c.:

\begin{eqnarray*}
U_{L} \left( x^{\mu}, -\phi, \rho \right) & = & 
U_{L} \left( x^{\mu}, \phi, \rho \right) \\
U_{R} \left( x^{\mu}, -\phi, \rho \right) & = & 
-U_{R} \left( x^{\mu}, \phi, \rho \right) \\
V_{L} \left( x^{\mu}, -\phi, \rho \right) & = & 
V_{L} \left( x^{\mu}, \phi, \rho \right) \\
V_{R} \left( x^{\mu}, -\phi, \rho \right) & = & 
-V_{R} \left( x^{\mu}, \phi, \rho \right) \\
U_{L} \left( x^{\mu}, \phi + \phi_{c}, \rho \right) & = & 
U_{L} \left( x^{\mu}, \phi_{c} -\phi, \rho \right) \\
U_{R} \left( x^{\mu}, \phi + \phi_{c}, \rho \right) & = & 
-U_{R} \left( x^{\mu}, \phi_{c} - \phi, \rho \right) \\
V_{L} \left( x^{\mu}, \phi + \phi_{c}, \rho \right) & = & 
V_{L} \left( x^{\mu}, \phi_{c} - \phi, \rho \right) \\
V_{R} \left( x^{\mu}, \phi + \phi_{c}, \rho \right) & = &
-V_{R} \left( x^{\mu}, \phi_{c} - \phi, \rho \right) \\
U_{L} \left( x^{\mu}, \phi, -\rho \right) & = &
V_{L} \left( x^{\mu}, \phi, \rho \right) \\
U_{R} \left( x^{\mu}, \phi, -\rho \right) & = &
V_{R} \left( x^{\mu}, \phi, \rho \right) \\
V_{L} \left( x^{\mu}, \phi, -\rho \right) & = & 
U_{L} \left( x^{\mu}, \phi, \rho \right) \\
V_{R} \left( x^{\mu}, \phi, -\rho \right) & = & 
U_{R} \left( x^{\mu}, \phi, \rho \right) \\
U_{L} \left( x^{\mu}, \phi, \rho + \rho_{c} \right) & = &
V_{L} \left( x^{\mu}, \phi, \rho_{c} - \rho \right) \\
U_{R} \left( x^{\mu}, \phi, \rho + \rho_{c} \right) & = &
V_{R} \left( x^{\mu}, \phi, \rho_{c} - \rho \right) \\
V_{L} \left( x^{\mu}, \phi, \rho_{c} + \rho \right) & = &
U_{L} \left( x^{\mu}, \phi, \rho_{c} - \rho \right) \\
V_{R} \left( x^{\mu}, \phi, \rho_{c} + \rho \right) & = &
U_{R} \left( x^{\mu}, \phi, \rho_{c} - \rho \right) \\
\end{eqnarray*}

Note that the representations
for the two $Z_{2}$ symmetries $\phi \rightarrow
-\phi$ and $\rho \rightarrow -\rho$ acting on
the fermions do not commute.
Because $\left[ S_{\phi}, S_{\rho} \right] \neq 0$,
we cannot find a decomposition of the fermion field
$\psi$ such that its components have definite
$Z_{2}$ transformation properties (even or odd)
under both $Z_{2}$'s.
This fact may seem odd, given our intuition
that these two $Z_{2}$ symmetries are independent.
In \cite{Hung:2001hw}, the fermion representation
of these $Z_{2}$'s do commute, and boundary
conditions are derived for component fields
with definite $Z_{2}$ transformation properties
under both symmetries.
As we demonstrate below, this treatment is
fundamentally not possible.
It follows that the action is not invariant under 
both symmetries in the inconsistent representation
given in \cite{Hung:2001hw}, as the reader
may verify.
With incorrect boundary conditions, the
resultant solutions are not trustworthy.
We now demonstrate why $S_{\phi}$ and $S_{\rho}$
must anticommute.

As is well-known, fermions have the odd
property of going into minus themselves when rotated
by $ 2 \pi$ (pun intended).
Applying successively the discrete transformations
$\phi \rightarrow -\phi$ followed by
$\rho \rightarrow -\rho$ is equivalent to
a rotation of $\pi$ radians in the $\phi - \rho$ 
plane.
Applying these discrete transformations
in the opposite order is equivalent to a rotation
of $-\pi$ radians in the same plane.
The difference in angles of these two rotations
is $2 \pi$.
Applying both transformations, first in the order
$S_{\rho} S_{\phi}$ and then in the order
$S_{\phi} S_{\rho}$, is equivalent to a rotation
of $2 \pi$ in the $\phi - \rho$ plane, and must
result in an overall minus sign for the fermion
field.
Therefore we must have $\left\{ S_{\phi}, S_{\rho} \right\} = 
0$ in order to be consistent with the spinor 
nature of fermions.

From these boundary conditions we see that
$U_{R} = V_{R} = 0$ on the entire boundary of rectangular region of 
dimensions $\phi_{c} \times \rho_{c}$ of our
fundamental cell.
It also follows that both $U_{L}$ and $V_{L}$ are even 
functions of the $\phi$ coordinate and that $U_{L} = V_{L}$ on
the $\rho = 0$ and $\rho = \rho_{c}$ boundaries.

\section{Scalar Field Zero Mode and Quark Mass Matrices}
In this section, we solve for the zero mode of a 
six dimensional Higgs scalar field.
As in the fermionic case, we also include a mass term for
the scalar in the six dimensional action.

We consider a real scalar field $\Phi$ propagating in a six
dimensional curved background described by the metric~(\ref{metric}).
Including a six dimensional mass term, the action is:

\begin{equation}
S = \frac{1}{2} \int \mbox{d}^{4}x \int \mbox{d} \phi \mbox{d} \rho 
\sqrt{G} \left( G^{AB} \partial_{A} \Phi \partial_{B} \Phi -
m_{\Phi}^{2} \Phi^{2} \right)
\end{equation}
Integrating by parts, one obtains:
\begin{eqnarray*}
S & = & \frac{1}{2} \int \mbox{d}^{4}x \int \mbox{d} \phi \mbox{d} \rho
\left( \frac{ e^{\sigma} e^{\gamma} }{ \left( e^{\sigma} + e^{\gamma}
 - 1 \right)^{4} } \eta^{\mu \nu} \partial_{\mu} \Phi \partial_{\nu}
\Phi + \Phi \partial_{\phi} \left( \frac{ e^{\gamma} }{
e^{\sigma} \left( e^{\sigma} + e^{\gamma} - 1 \right)^{4} } \partial_{\phi}
 \Phi \right) \right. \\
& & + \left. \Phi \partial_{\rho} \left( \frac{ e^{\sigma} }{ e^{\gamma}
\left( e^{\sigma} + e^{\gamma} - 1 \right)^{4} } \partial_{\rho}
 \Phi \right) - m_{\Phi}^{2} \Phi^{2} \sqrt{G} \right) 
\end{eqnarray*}

In order to give a four dimensional interpretation to this action, we
go through the dimensional reduction procedure. 
Thus we decompose the six dimensional field into KK modes
\begin{equation}
\Phi \left( x, \phi, \rho \right) = \sum_{n,m} \phi_{n,m} \left( x
 \right) f_{n,m} \left( \phi, \rho \right)
\end{equation}
Using this decomposition, the above action becomes a sum over
KK modes.

\begin{equation}
S = \frac{1}{2} \sum_{n,m} \int \mbox{d}^{4}x \left\{ \eta^{\mu \nu}
 \partial_{\mu} \phi_{n,m} \left( x \right) \partial_{\nu}
\phi_{n,m} \left( x \right) - m_{n,m}^{2} \phi_{n,m}^{2} \left( x
 \right) \right\}
\end{equation}

In order to reproduce the canonical four dimensional
kinetic terms,
we need to impose the orthogonality relations:
\begin{equation}
\int \int \mbox{d} \phi \mbox{d} \rho \frac{ e^{\sigma} e^{\gamma} }{
 \left( e^{\sigma} + e^{\gamma} - 1 \right)^{4} } f_{m,i}^{\ast} \left(
\phi, \rho \right) f_{n,k} \left( \phi, \rho \right) = \delta_{mn},
\delta_{ik}
\end{equation}
Varying the action with respect to $f_{n}$ leads to
the following equation for the zero mode:
\begin{equation}
\partial_{\phi} \left( \frac{ e^{\gamma} }{ e^{\sigma} \left(
e^{\sigma} + e^{\gamma} - 1 \right)^{4} } \partial_{\phi} f_{0} \right)
  +  \partial_{\rho} \left( \frac{ e^{\sigma} }{ e^{\gamma} \left(
e^{\sigma} + e^{\gamma} - 1 \right)^{4} } \partial_{\rho} f_{0} \right)
 -\frac{ m^{2} e^{\sigma} e^{\gamma} }{ \left( e^{\sigma}
+ e^{\gamma} - 1 \right)^{6} } f_{0} = 0
\end{equation}
From the form of this equation, we expect the 
$\phi$ and $\rho$ dependence of $f_{0}$ to be the same.
Hence, we write

\begin{eqnarray*}
\partial_{\phi} \left( \frac{ e^{\gamma} }{ e^{\sigma} \left(
e^{\sigma} + e^{\gamma} - 1 \right)^{4} } \partial_{\phi} f_{0} \right)
& = & 
\frac{ m^{2} e^{\sigma} e^{\gamma} }{ 2 \left( e^{\sigma}
+ e^{\gamma} - 1 \right)^{6} } f_{0} \\
   \partial_{\rho} \left( \frac{ e^{\sigma} }{ e^{\gamma} \left(
e^{\sigma} + e^{\gamma} - 1 \right)^{4} } \partial_{\rho} f_{0} \right)
& = & 
\frac{ m^{2} e^{\sigma} e^{\gamma} }{ 2 \left( e^{\sigma}
+ e^{\gamma} - 1 \right)^{6} } f_{0}
\end{eqnarray*}

Any solution to this pair of equations automatically
satisfies the original equation.

The first of the above equations can be written as

\begin{equation}
\left( \frac{ 4 \sigma^{\prime} e^{\gamma} }{ \left( e^{\sigma}
+ e^{\gamma} - 1 \right) } + \frac{ \sigma^{\prime} e^{\gamma} }{
 e^{\sigma} } \right) \partial_{\phi} f_{0}  - 
\frac{ e^{\gamma} }{ e^{\sigma} } \partial_{\phi}^{2} f_{0} -
\frac{ m^{2} e^{\sigma} e^{\gamma} f_{0} }{ 2 \left(
e^{\sigma} + e^{\gamma} - 1 \right)^{2} } = 0
\end{equation}

Substituting in the forms for $\sigma$ and $\gamma$, we arrive at
the two equations:
\begin{equation}
\left( \frac{ 4 k e^{ k \rho} }{ \left( e^{k \phi}
+ e^{ k \rho} - 1 \right) } + \frac{ k e^{k \rho} }{
 e^{ k \phi} } \right) \partial_{\phi} f_{0}  - 
\frac{ e^{k \rho} }{ e^{k \phi} } \partial_{\phi}^{2} f_{0} -
\frac{ m^{2} e^{k \phi} e^{k \rho} f_{0} }{ 2 \left(
e^{k \phi} + e^{k \rho} - 1 \right)^{2} } = 0
\end{equation}

\begin{equation}
\left( \frac{ 4 k e^{ k \sigma} }{ \left( e^{k \phi}
+ e^{ k \rho} - 1 \right) } + \frac{ k e^{k \sigma} }{
 e^{ k \rho} } \right) \partial_{\rho} f_{0}  - 
\frac{ e^{k \phi} }{ e^{k \rho} } \partial_{\rho}^{2} f_{0} -
\frac{ m^{2} e^{k \phi} e^{k \rho} f_{0} }{ 2 \left(
e^{k \phi} + e^{k \rho} - 1 \right)^{2} } = 0
\end{equation}

 For the special case when 

\begin{equation}
2 k^{2} m^{2} = 25 k^{4}
\end{equation}

we can find an analytic solution:

\begin{equation}
f \left( \phi, \rho \right) = e^{ \left( \frac{ \ln \left( e^{k \phi} + 
e^{k \rho} - 1 \right) }{ 2 k^{2} } \right) }
\end{equation}
 

\vglue 5cm
\hglue 5.5cm
\psfig{figure=plot.ps,height=8cm,angle=0}
\begin{quote}
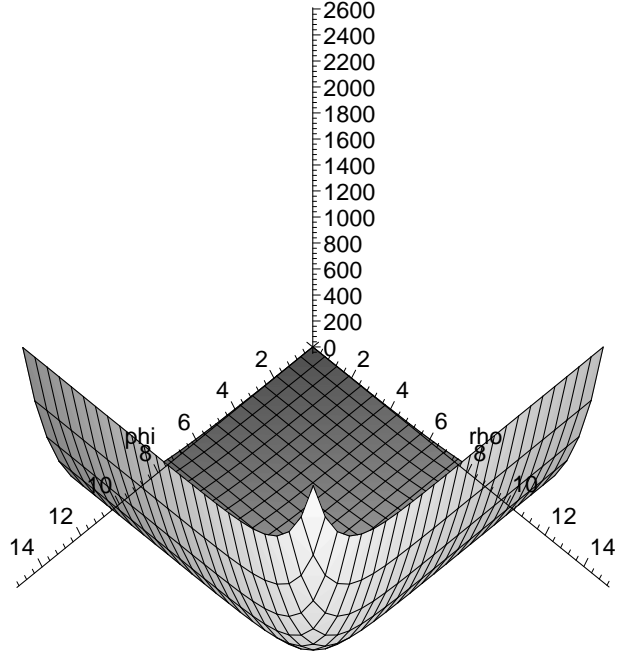
\scriptsize Fig. (4): A plot of the
 scalar profile for the case $k=1$.
\end{quote}

In Fig. (4) we plot the scalar profile for the case $k=1$.

Having presented the equations for the fermion
field profiles in the extra dimensions
and having demonstrated the complex nature of the solutions,
we may now see how CP violation emerges naturally.
From the fermions and the 
Higgs scalar, we construct
a six dimensional lorentz invariant Yukawa interaction from a term
such as
\begin{equation}
\sqrt{-G} \lambda_{ij}^{(6)} H \overline{\psi_{i}} \psi_{j}
\end{equation}
where the fermion fields are eight component objects and their
indices $i$ and $j$ are generation labels with the 
Yukawa coupling matrices $\lambda_{ij}^{(6)}$ 
connecting only fermions in combinations consistent with
gauge group invariance.
The fermion bilinear may be expressed as
\begin{eqnarray*}
\overline{\psi} \psi & =  &
-i \overline{\psi_{-}} \psi_{+} + i 
\overline{\psi_{+}} \psi_{-} \\
 & = & i \overline{\psi_{L+}} \psi_{R-} +
 i \overline{\psi_{R+}} \psi_{L-} \\
 & & -i \overline{\psi_{L-}} \psi_{R+} - i
 \overline{\psi_{R-}} \psi_{L+}
\end{eqnarray*}
where the conjugate fields are formed with the usual $\gamma_{0}$
of the four dimensional Dirac algebra.

One may write the exact same form of the Yukawa
coupling in the six dimensional action as in the
SM action:
\begin{equation}
\int \mbox{d}^{4} x \int \mbox{d} \phi \mbox{d} \rho
\sqrt{-G} \lambda_{ij}^{(6)} H \overline{\psi_{i}} \psi_{j}
\end{equation}
Recall, however, that the motivation for considering the 
implications of extra dimensions is to see if their existence
can help to reduce or simplify 
the redundancy inherent in the four dimensional Yukawa coupling
matrices.
Allowing $\lambda_{ij}^{(6)}$ and the six dimensional fermion mass
terms $m_{i}$ to be arbitrary parameters only 
increases the redundancy of physical information contained 
in the parameters of the action, and, from the
perspective of attempting to gain deeper understanding
of the fermion mass hierarchy, considerably weakens the
motivation of considering extra dimensions.
However, one
starting point often adopted in attempts to study
the fermion mass hierarchy is to start with a discrete
flavor symmetry which leads to the so-called democratic mass
matrix, with all entries being the same.
This simplest ans\"{a}tz leads to one massive eigenstate
and two degenerate massless eigenstates and is thus
considered a reasonably successful approximation
given the simplicity of the ans\"{a}tz.
In \cite{Hung:2001hw}, the introduction of a single flat extra 
dimension provides a theoretical explanation of the
democratic ans\"{a}tz itself in terms of higher
dimensional geography rather than some additional
flavor physics.
The introduction of an additional flat extra dimension and
its associated Dirac algebra then allow for the 
possibility to realize a more realistic spectrum than that 
provided by the democratic mass matrices.
Because the extra dimensions are taken to be flat, the
fermion profiles may be separable
functions of each of the extra dimensions.
This property transforms the effective
democratic quark mass matrices one 
obtains from dimensionally reducing from five to four
dimensions into pure phase mass matrices.

The success of this approach in the flat space scenario 
cannot be carried over without modification to the case
of two warped extra dimensions.
Because the equations for the fermion profiles in
the warped scenario we are considering
are not separable,
it is not possible to automatically achieve pure
phase effective four dimensional quark mass matrices within
our context.

If we set $\lambda_{ij}^{(6)}$ equal to the democratic
mass matrix in both the up and down quark sectors, we
arrive at essentially the same level of understanding as
when the democratic form is adopted in flavor
symmetry approaches to the problem in four dimensions.
In that case, a longstanding difficulty has been to find
a successful implementation of a breaking of this symmetry,
(sometimes involving the additional physics of a ``flavon''
field) with the additional complication of 
ultimately arriving at complex mass matrices.

The principle observation of this work is that the presence 
of the extra dimensions serves to break the 
democratic form of the $\lambda_{ij}^{(6)}$, which is now
 controlled by the
six dimensional fermion mass terms down to the 
effective $\lambda_{ij}^{(4)}$.
Geography in the extra warped dimensions is an
alternative to the flavon field method of breaking the 
flavor symmetry of the Yukawa terms.
In addition, this geometric method of breaking the 
flavor symmetry naturally leads to complex 
effective four dimensional mass matrices.

Adopting the democratic form for $\lambda_{ij}^{(6)}$ in
each quark sector does not result in a calculable
model of flavor mixing and masses.
It does, however, correspond to a so-called
minimal parameter basis.
Six quark masses and four flavor mixing parameters are
derived from ten six-dimensional parameters.
This minimal parameter set arises from the 
following considerations.
The SM gauge symmetries allow for the existence of nine
different six-dimensional fermion masses $m_{i}$.
The left-handed up-type and down-type quarks of the same
generation must have the same six-dimensional mass
parameter $m$ because together they form an $SU \left( 2 \right)$
doublet.
The right-handed components of the up and down type
quarks for each generation have different mass parameters.
From the four dimensional perspective, a massive fermion 
field is formed only after electroweak symmetry breaking.
In addition to these nine independent mass parameters,
we also have the freedom associated with the 
curvature $k$ of the bulk AdS space.
We are assuming equal magnitudes for all the 4-brane
tensions and so have only one dimension-full parameter $k$, instead
of two as would be the case if we allowed arbitrary 
4-brane tensions.
Requiring that this setup reproduce the RS resolution 
of the gauge hierarchy problem then fixes the coordinate
length of the fundamental cell we are considering.

\section{Conclusion}

We first remind the reader of some of the 
promising results already attained in addressing the fermion
mass and mixing hierarchy problems within the context
of extra dimensions \cite{Hung:2001hw,Huber:2000ie,Arkani-Hamed:1999dc,Mouslopoulos:2001uc}.
Considering the $2 \times 2$ matrix with columns labelled
``one extra dimension'' and ``two extra dimensions'' and
rows labelled ``flat extra space'' and ``warped extra space'', 
we have shown that the $( 2, 2 )$ element is also a possible
arena in which to study the problem.
But the principle motivations for considering the
two warped extra dimensions scenario are not just 
for the sake of completeness.
The possibility of attaining the CP violating phase in the
quark flavor-mixing matrix via this higher dimensional
mechanism is a property of the Dirac algebra in six
dimensions and remains a possibility whether or not the two
extra dimensions are flat or warped.
Independent of considerations of the gauge hierarchy
problem, one advantage of the warped case over the flat 
case is that no
additional physics is required in order to
localize the fermions.
In the flat space case, the localization mechanism
of the fermions involves the introduction of a scalar field and
 specific forms for the scalar field must be assumed. 
 Even 
then the fermion profiles are solved analytically only after making 
approximations to the given scalar field \cite{Arkani-Hamed:1999dc}.

We have shown in this work that CP violation may be understood
as a natural consequence of the Dirac algebra in 
six dimensions.
Another motivation for going to six dimensions
concerns the mystery of the generation index.
As shown in \cite{Kogan:2001wp}, multibrane world scenarios imply the 
existence of light KK modes that are suggestive of
family replication.
Going to six dimensions may alleviate some of the 
difficulties encountered in trying to implement this
program.
This possibility is currently under investigation.

\section*{Acknowledgements}
We would like to thank L.~Brown, C.~Burgess, J.~Cline, P.~Hung and M.~Seco
for useful discussions.  D.D. and D.E. are supported in part
by NSERC (Canada) and FCAR (Quebec). K.K. is supported by
the US Department of Energy under Contract DE-FG0291ER40688, TASK A.

\bibliography{reff}

\begin{thebibliography}{10}

\bibitem{Weinberg:1977hb}
Steven Weinberg.
\newblock The problem of mass.
\newblock {\em Trans. New York Acad. Sci.}, 38:185--201, 1977.

\bibitem{Wilczek:1977uh}
Frank Wilczek and A.~Zee.
\newblock Discrete flavor symmetries and formula for the cabibbo angle.
\newblock {\em Phys. Lett.}, B70:418, 1977.

\bibitem{Rothman:1979ft}
Arthur~C. Rothman and Kyungsik Kang.
\newblock Calculability: The mass matrix and cp violation.
\newblock {\em Phys. Rev. Lett.}, 43:1548, 1979.

\bibitem{Kang:1981yg}
Kyungsik Kang and Arthur~C. Rothman.
\newblock Generalized mixing angles in gauge theories with natural flavor
  conservation.
\newblock {\em Phys. Rev.}, D24:167, 1981.

\bibitem{Fritzsch:1978vd}
H.~Fritzsch.
\newblock Weak interaction mixing in the six - quark theory.
\newblock {\em Phys. Lett.}, B73:317--322, 1978.

\bibitem{Fritzsch:1979zq}
Harald Fritzsch.
\newblock Quark masses and flavor mixing.
\newblock {\em Nucl. Phys.}, B155:189, 1979.

\bibitem{DeRujula:1977ry}
A.~De Rujula, Howard Georgi, and S.~L. Glashow.
\newblock A theory of flavor mixing.
\newblock {\em Ann. Phys.}, 109:258, 1977.

\bibitem{Georgi:1979dq}
Howard Georgi and D.~V. Nanopoulos.
\newblock Ordinary predictions from grand principles: t quark mass in o(10).
\newblock {\em Nucl. Phys.}, B155:52, 1979.

\bibitem{Kang:1997uv}
Kyungsik Kang and Sin~Kyu Kang.
\newblock New class of quark mass matrix and calculability of flavor mixing
  matrix.
\newblock {\em Phys. Rev.}, D56:1511--1514, 1997.

\bibitem{Thirring:1972de}
W.~E. Thirring.
\newblock Fivedimensional theories and cp-violation.
\newblock {\em Acta Phys. Austriaca Suppl.}, 9:256--271, 1972.

\bibitem{Casadio:2001fe}
R.~Casadio and A.~Gruppuso.
\newblock Discrete symmetries and localization in a brane-world.
\newblock {\em Phys. Rev.}, D64:025020, 2001.

\bibitem{Chang:2001uk}
Darwin Chang, Wai-Yee Keung, and Rabindra~N. Mohapatra.
\newblock Models for geometric cp violation with extra dimensions.
\newblock {\em Phys. Lett.}, B515:431--441, 2001.

\bibitem{Chang:2001yn}
Darwin Chang and R.~N. Mohapatra.
\newblock Geometric cp violation with extra dimensions.
\newblock {\em Phys. Rev. Lett.}, 87:211601, 2001.

\bibitem{Huang:2001np}
Chao-Shang Huang, Tian-jun Li, Wei Liao, and Qi-Shu Yan.
\newblock Cp violation and extra dimensions. \it hep-ph/0101002, \rm
\newblock 2001.

\bibitem{Branco:2000rb}
G.~C. Branco, Andre de~Gouvea, and M.~N. Rebelo.
\newblock Split fermions in extra dimensions and cp violation.
\newblock {\em Phys. Lett.}, B506:115--122, 2001.

\bibitem{Sakamura:2000ik}
Yutaka Sakamura.
\newblock Our wall as the origin of cp violation.
\newblock{\em hep-th/0011098}, 2000.

\bibitem{Sakamura:1999fa}
Yutaka Sakamura.
\newblock Spontaneous cp violation in large extra dimensions.
\newblock {\em hep-ph/9909454},  1999.

\bibitem{Hung:2001hw}
P.~Q. Hung and M.~Seco.
\newblock Pure phase mass matrices from six dimensions.
\newblock {\em hep-ph/0111013}, 2001.

\bibitem{Huber:2000ie}
Stephan~J. Huber and Qaisar Shafi.
\newblock Fermion masses, mixings and proton decay in a randall- sundrum model.
\newblock {\em Phys. Lett.}, B498:256--262, 2001.

\bibitem{Arkani-Hamed:1999dc}
Nima Arkani-Hamed and Martin Schmaltz.
\newblock Hierarchies without symmetries from extra dimensions.
\newblock {\em Phys. Rev.}, D61:033005, 2000.

\bibitem{Mouslopoulos:2001uc}
Stavros Mouslopoulos.
\newblock Bulk fermions in multi-brane worlds.
\newblock {\em JHEP}, 05:038, 2001.

\bibitem{Kaplanet:2001}
David E. Kaplan and Tim M.P. Tait
\newblock New tools for fermion masses from extra dimensions.
\newblock {\em JHEP}, 051:0111, 2001.

\bibitem{Chodos:1999zt}
Alan Chodos and Erich Poppitz.
\newblock Warp factors and extended sources in two transverse dimensions.
\newblock {\em Phys. Lett.}, B471:119--127, 1999.

\bibitem{Gherghettaet:2000}
Tony Gherghetta and Mikhail Shaposhnikov.
\newblock Localizing gravity on a string-like defect in six dimensions.
\newblock {\em Phys. Rev. Lett.}, 85,  240-243, 2000.

\bibitem{Collins:2001ni}
Hael Collins and Bob Holdom.
\newblock The randall-sundrum scenario with an extra warped dimension.
\newblock {\em Phys. Rev.}, D64:064003, 2001.

\bibitem{Kogan:2001yr}
Ian~I. Kogan, Stavros Mouslopoulos, Antonios Papazoglou, and Graham~G. Ross.
\newblock Multigravity in six dimensions: Generating bounces with flat positive
  tension branes.
\newblock {\em Phys. Rev.}, D64:124014, 2001.

\bibitem{Burgess:2001bn}
C.~P. Burgess, James~M. Cline, Neil~R. Constable, and Hassan Firouzjahi.
\newblock Dynamical stability of six-dimensional warped brane- worlds.
\newblock {\em JHEP} 0201:014, 2002.

\bibitem{Kim:2001rm}
Jihn~E. Kim, Bumseok Kyae, and Hyun~Min Lee.
\newblock Localized gravity and mass hierarchy in d = 6 with gauss- bonnet
  term.
\newblock {\em Phys. Rev.}, D64:065011, 2001.

\bibitem{Arkani-Hamed:1998rs}
Nima Arkani-Hamed, Savas Dimopoulos, and G.~R. Dvali.
\newblock The hierarchy problem and new dimensions at a millimeter.
\newblock {\em Phys. Lett.}, B429:263--272, 1998.

\bibitem{Arkani-Hamed:1998nn}
Nima Arkani-Hamed, Savas Dimopoulos, and G.~R. Dvali.
\newblock Phenomenology, astrophysics and cosmology of theories with
  sub-millimeter dimensions and tev scale quantum gravity.
\newblock {\em Phys. Rev.}, D59:086004, 1999.

\bibitem{Antoniadis:1998ig}
Ignatios Antoniadis, Nima Arkani-Hamed, Savas Dimopoulos, and G.~R. Dvali.
\newblock New dimensions at a millimeter to a fermi and superstrings at a tev.
\newblock {\em Phys. Lett.}, B436:257--263, 1998.

\bibitem{Gogberashvili:1998vx}
Merab Gogberashvili.
\newblock Hierarchy problem in the shell-universe model.
\newblock {\em hep-ph/9812296}, 1998.

\bibitem{Randall:1999ee}
Lisa Randall and Raman Sundrum.
\newblock A large mass hierarchy from a small extra dimension.
\newblock {\em Phys. Rev. Lett.}, 83:3370--3373, 1999.

\bibitem{Randall:1999vf}
Lisa Randall and Raman Sundrum.
\newblock An alternative to compactification.
\newblock {\em Phys. Rev. Lett.}, 83:4690--4693, 1999.

\bibitem{Chang:1999nh}
Sanghyeon Chang, Junji Hisano, Hiroaki Nakano, Nobuchika Okada, and Masahiro
  Yamaguchi.
\newblock Bulk standard model in the randall-sundrum background.
\newblock {\em Phys. Rev.}, D62:084025, 2000.

\bibitem{Kogan:2001wp}
Ian~I. Kogan, Stavros Mouslopoulos, Antonios Papazoglou, and Graham~G. Ross.
\newblock Multi-localization in multi-brane worlds.
\newblock {\em Nucl. Phys.}, B615:191--218, 2001.

\bibitem{Hatanaka:1999ac}
Hisaki Hatanaka, Makoto Sakamoto, Motoi Tachibana, and Kazunori Takenaga.
\newblock Many-brane extension of the randall-sundrum solution.
\newblock {\em Prog. Theor. Phys.}, 102:1213--1218, 1999.

\bibitem{Grossman:1999ra}
Yuval Grossman and Matthias Neubert.
\newblock Neutrino masses and mixings in non-factorizable geometry.
\newblock {\em Phys. Lett.}, B474:361--371, 2000.

\end{thebibliography}
\end{document}